\documentclass[a4paper]{article}
\usepackage{enumitem,hyperref,dahlin,natbib,multirow,rotating,sectsty,booktabs}           
\allsectionsfont{\sffamily}
\usepackage[margin=1in]{geometry}
\usepackage{setspace}

\newtheorem{definition}{Definition}

\newtheorem{proposition}{Proposition}

\title{Quasi-Newton particle Metropolis-Hastings} 

\author{Johan Dahlin\thanks{Department of Electrical Engineering, Link{\"o}ping University, Link{\"o}ping, SE. E-mail: \textit{johan.dahlin@liu.se}.}
, Fredrik Lindsten\thanks{Department of Engineering, University of Cambridge, Cambridge, UK. E-mail: \textit{fredrik.lindsten@eng.cam.ac.uk}.} \,
and Thomas B.\ Sch\"{o}n\thanks{Department of Information Technology, Uppsala University, Uppsala, SE. E-mail: \textit{thomas.schon@it.uu.se}.}
}

\begin{document}
\maketitle
\doublespacing

\begin{abstract}
\noindent Particle Metropolis-Hastings enables Bayesian parameter inference in general nonlinear state space models (SSMs). However, in many implementations a random walk proposal is used and this can result in poor mixing if not tuned correctly using tedious pilot runs. Therefore, we consider a new proposal inspired by quasi-Newton algorithms that may achieve similar (or better) mixing with less tuning. An advantage compared to other Hessian based proposals, is that it only requires estimates of the gradient of the log-posterior. A possible application is parameter inference in the challenging class of SSMs with intractable likelihoods. We exemplify this application and the benefits of the new proposal by modelling log-returns of future contracts on coffee by a stochastic volatility model with $\alpha$-stable observations. 
\end{abstract}

\newpage
\section{Introduction}
\label{sec:intro}
% Introduce the state space model
We are interested in Bayesian parameter inference in the nonlinear state space model (SSM) possibly with an intractable likelihood. An SSM with latent states $x_{0:T}=\{x_t\}_{t=0}^T$ and observations $y_{1:T}$ is given by
\begin{align}
	x_{t+1}|x_t \sim f_{\theta}(x_{t+1}|x_{t}), \qquad
	y_{t}  |x_t \sim g_{\theta}(y_{t  }|x_{t}),
\label{eq:SSMdef}
\end{align}
with $x_0 \sim \mu_{\theta}(x_0)$ and where $\theta \in \Theta \subseteq \mathbb{R}^p$ denotes the static unknown parameters. Here, we assume that it is possible to simulate from the distributions $\mu_{\theta}(x_0)$, $f_{\theta}(x_{t+1}|x_{t})$ and $g_{\theta}(y_{t}|x_{t})$, even if the respective densities are unavailable.

% Discuss what we mean with an intractable likelihood and why it appears (no expression or expensive to evaluate)
The main object of interest in Bayesian parameter inference is the \textit{parameter posterior} distribution,
\begin{align}
	\pi(\theta) = p(\theta|y_{1:T}) \propto p_{\theta}(y_{1:T}) p(\theta),
	\label{eq:parameterposterior}
\end{align}	 
which is often intractable and cannot be computed in closed form. The problem lies in that the likelihood $p_{\theta}(y_{1:T}) = p(y_{1:T}|\theta)$ cannot be exactly computed. However, it can be estimated by computational statistical methods such as sequential Monte Carlo (SMC; \citealp{DoucetJohansen2011}). The problem is further complicated when  $g_{\theta}(y_t|x_t)$ cannot be evaluated point-wise, which prohibits direct application of SMC. This could be the result of that the density does not exist or that it is computationally prohibitive to evaluate. In both cases, we say that the likelihood of the SSM \eqref{eq:SSMdef} is \textit{intractable}.

%\begin{align} 
%	p_{\theta}(y_{1:T}) = p(y_{1:T}|\theta) = p_{\theta}(y_1) \prod_{t=2}^T p_{\theta}(y_t|y_{1:t-1}),
%	\label{eq:likelihoodSSM}
%\end{align}

% \item Discuss ABC for doing inference in models with intractable likelihoods. \citep{MarinPudloRobertRyder2012}
Recent efforts to develop methods for inference in models with intractable likelihoods have focused on approximate Bayesian computations (ABC; \citealp{MarinPudloRobertRyder2012}). The main idea in ABC is that data \textit{simulated} from the model (using the correct parameters) should be \textit{similar} to the observed data. This idea can easily be incorporated into many existing inference algorithms, see \citet{DeanSinghJasraPeters2014}.

% SMC-ABC can be used to estimate the likelihood unbiasedly and therefore PMH can be used. \citep{JasraSinghMartinMcCoy2012}. Poor mixing in the marginal PMH-ABC algorithm.
An example of this is the ABC version of particle Metropolis-Hastings (PMH-ABC; \citealp{Jasra2015,BornnPillaiSmithWoordward2014}). In this algorithm, the intractable likelihood is replaced with an estimate obtained by the ABC version of SMC (SMC-ABC; \citealp{JasraSinghMartinMcCoy2012}). However, the random walk proposal often used in PMH(-ABC) can result in problems with poor mixing, which leads to high variance in the posterior estimates. The mixing can be improved by \textit{pre-conditioning} the proposal with a matrix $\mathcal{P}$, which typically is chosen as the unknown posterior covariance \citep{SherlockThieryRobetsRosenthal2015}. However, estimating $\mathcal{P}$ can be challenging when $p$ is large or the posterior \eqref{eq:parameterposterior} is non-isotropic. This typically results in that the user needs to run many tedious pilot runs when implementing PMH(-ABC) for parameter inference in a new SSM.

% Main contribution is to extend the framework from \citep{DahlinLindstenSchon2014c} to models with intractable likelihoods.
Our main contribution is to adapt a \textit{limited-memory BFGS algorithm} \citep{NocedalWright2006} as a proposal in PMH(-ABC). This is based on earlier work by \cite{DahlinLindstenSchon2014c} and \cite{ZhangSutton2011}. In the former, we discuss how to make use of gradient ascent and Newton-type proposals in PMH. The advantages of the new proposal are; (i) good mixing when gradient estimates are accurate, (ii) no tedious pilot runs required and (iii) only requires gradients to approximate the local Hessian. These advantages are important as direct estimation of gradients using SMC often is simpler than for Hessians. Note that this new proposal is useful both with and without the ABC approximation of the likelihood.

To demonstrate these benefits we consider a linear Gaussian state space (LGSS) model, where we compare the performance of our proposal with and without ABC. Furthermore, we consider using a stochastic volatility model with $\alpha$-stable log-returns \citep{Nolan2003} to model future contracts on coffee. This model is common in the ABC literature as the likelihood is intractable, see \citet{DahlinVillaniSchon2015}, \citet{Jasra2015} and \citet{YildirimSinghDeanJasra2014}.

\section{Particle Metropolis-Hastings}
\label{sec:overview}
%\item PMH \citep{AndrieuDoucetHolenstein2010} is an MCMC method for estimating the parameter posterior distribution.
A popular approach to estimate the parameter posterior~\eqref{eq:parameterposterior} is to make use of statistical simulation methods. PMH \citep{AndrieuDoucetHolenstein2010} is one such method and it operates by constructing a Markov chain, which has the sought posterior as its stationary distribution. As a result, we obtain samples from the posterior by simulating the Markov chain to convergence.

%\item The construction of the chain depends on a proposal distribution and point-wise evaluations of the posterior distribution.
The Markov chain targeting \eqref{eq:parameterposterior} is constructed by an iterative procedure. During iteration $k$, we propose a candidate parameter $\theta' \sim q(\theta'|\theta_{k-1},u_{k-1})$ and an \textit{auxiliary variable} $u' \sim m_{\theta'}(u')$ as detailed in the following using proposals $q$ and $m_{\theta'}$. The candidate is then \textit{accepted}, i.e.\ $\{\theta_k , u_k \} \leftarrow \{\theta', u'\}$, with the probability
\begin{align}
	\alpha \big( \theta',\theta_{k-1},u',u_{k-1} \big) 
	= 
	\frac{\widehat{\pi} \big( \theta' \big| u' \big)}{\widehat{\pi} \big( \theta_{k-1} \big| u_{k-1} \big)}
	\frac{q \big( \theta_{k-1} \big| \theta',u' \big) }{q \big( \theta' \big| \theta_{k-1},u_{k-1} \big) },
	\label{eq:PMHaprob}
\end{align}
otherwise the parameter is \textit{rejected}, i.e.\ $\{\theta_k, u_k \} \leftarrow \{ \theta_{k-1}, u_{k-1}\}$. Here, $\widehat{\pi}(\theta|u)=\widehat{p}_{\theta}(y_{1:T}|u) p(\theta)$ denotes some \textit{unbiased} estimate of $\pi(\theta)$ constructed using $u$ and $p(\theta)$ denotes the parameter prior distribution.

PMH can be viewed as a Metropolis-Hastings algorithm in which the intractable likelihood is replaced with an unbiased noisy estimate. It is possible to show that this so-called \textit{exact approximation} results in a valid algorithm as discussed by \citet{AndrieuRoberts2009}. Specifically, the Markov chain generated by PMH converges to the desired stationary distribution despite the fact that we are using an approximation of the likelihood. It is also possible to show that $u$ can be included into the proposal $q$, which is necessary for including gradients and Hessians when proposing $\theta'$ as discussed by \cite{DahlinLindstenSchon2014c}.

%\item Give the general expression for the algorithm.
%\item Discuss how IF relates to the variance of the estimate of a functional.
In Section~\ref{sec:abc}, we discuss how to construct the proposal~$m_{\theta}$ by running an SMC algorithm. In this case, the auxiliary variable~$u$ is the resulting generated particle system. We obtain PMH-ABC as presented in Algorithm~\ref{alg:pmh}, when SMC-ABC is used for Step~4. This is a complete procedure for generating $K$ correlated samples $\{\theta_{1},\ldots,\theta_{K}\}$ from \eqref{eq:parameterposterior}. By the ergodic theorem, we can estimate any posterior expectation of an integrable \textit{test function} $\varphi: \Theta \rightarrow \mathbb{R}$ (e.g.\ the posterior mean) by
\begin{align}
	\mathbb{E} \Big[ \varphi(\theta) \big| y_{1:T} \Big] 
	%= 
	%\dint \varphi(\theta) p(\theta|y_{1:T}) \dd \theta 
	\approx 
	\widehat{\varphi}_{\text{MH}}
	\triangleq	
	\frac{1}{K - K_b} \sum_{k=K_b}^K \varphi(\theta_k)
	,
	\label{eq:functionalPosterior}
\end{align}
which is a strongly consistent estimator if the Markov chain is ergodic \citep{MeynTweedie2009}. Here, we discard the first $K_b$ samples known as the \textit{burn-in}, i.e.\ before the chain reaches stationarity. Under geometric mixing conditions, the error of the estimate obeys the central limit theorem (when $(K-K_b) \rightarrow \infty$) given by
\begin{align}
	\sqrt{K-K_b} \Big[ \widehat{\varphi}_{\text{MH}} - \mathbb{E} \big[ \varphi(\theta) \big| y_{1:T} \big] \Big]
	\stackrel{d}{\longrightarrow}
	\mathcal{N} \Big( 0,\sigma^2_{\varphi} \Big),
	\label{eq:functionalPosteriorCLT}
\end{align}
where $\sigma^2_{\varphi}$ denotes the variance of the estimator. The variance is proportional to the inefficiency factor (IF), which describes the \textit{mixing} of the Markov chain. Hence, we can use IF in the illustrations presented in Section~\ref{sec:results} to compare the mixing between different proposals.

\begin{algorithm}[!t]
\caption{\textsf{Particle Metropolis-Hastings (PMH)}}
\textsc{Inputs:} $K>0$ (no.\ MCMC steps), $\theta_0$ (initial parameters) \\ and $\{q,m_{\theta}\}$ (proposals). \\
\textsc{Output:} $\{\theta_1,\ldots,\theta_K\}$ (approximate samples from the posterior).
\algrule[.4pt]
\begin{algorithmic}[1]
	\STATE Generate $u_0 \sim m_{\theta_0}$ and compute $\widehat{p}_{\theta_0}(y_{1:T}|u_0)$.
	\FOR{$k=1$ to $K$}
		\STATE Sample $\theta' \sim q(\theta'|\theta_{k-1},u_{k-1})$.
		\STATE Sample $u' \sim m_{\theta'}$ using Algorithm~\ref{alg:smc}.
		\STATE Compute $\widehat{p}_{\theta'}(y_{1:T}|u')$ using \eqref{eq:likeEst}.
		\STATE Sample $\omega_k$ uniformly over $[0,1]$.
		\IF{$\omega_k \leq \min\{1,\alpha(\theta',\theta_{k-1},u',u_{k-1})\}$ given by \eqref{eq:PMHaprob}}
			\STATE Accept $\theta'$, i.e.\
			$\{\theta_k, u_k\} \leftarrow \{\theta',u'\}$. 
		\ELSE
			\STATE 
			Reject $\theta'$, i.e.\
			$\{\theta_k, u_k\} \leftarrow \{\theta_{k-1},u_{k-1}\}$.
		\ENDIF
	\ENDFOR
\end{algorithmic}
\label{alg:pmh}
\end{algorithm}

\section{Proposal for parameters}
\label{sec:proposals}
To complete Algorithm~\ref{alg:pmh}, we need to specify a proposal $q$ from which we sample $\theta'$. The choice of proposal is important as it is one of the factors that influences the mixing of resulting Markov chain. The general form of a Gaussian proposal discussed in \citet{DahlinLindstenSchon2014c} is
\begin{align}
	q \big( \theta''|\theta',u' \big)
	=
	\mathcal{N}
	\Big(
	\theta''; 
	\mu \big( \theta',u' \big),
	\Sigma \big( \theta',u' \big)
	\Big),
	\label{eq:GeneralPMHProposal}
\end{align}
where different choices of the mean function $\mu(\theta',u')$ and covariance function $\Sigma(\theta',u')$ results in different versions of PMH as presented in Table~\ref{tbl:pmh-proposals}.

\begin{table}[t]
\caption{Different proposals for the PMH algorithm.}
\begin{center}
\begin{tabular}{lcc}
\toprule
Proposal & $\mu(\theta',u')$ & $\Sigma(\theta',u')$ \\
\midrule
PMH0 
& 
$\theta'$
&
$\epsilon_0^2 \, \mathcal{P}^{-1}$
\\
PMH1 & 
$\theta' + \frac{\epsilon_1^2}{2} \big[ \mathcal{P}^{-1} \, \widehat{\mathcal{G}}(\theta'|u') \big]$ 
&
$\epsilon_1^2 \, \mathcal{P}^{-1}$
\\
PMH2 & 
$\theta' + \frac{\epsilon_2^2}{2} \big[ \widehat{\mathcal{H}}(\theta'|u') \big]^{-1} \, \widehat{\mathcal{G}}(\theta'|u')$
&
$\epsilon_2^2 \big[ \widehat{\mathcal{H}}(\theta'|u') \big]^{-1}$
\\
\bottomrule
\end{tabular}
\end{center}
\label{tbl:pmh-proposals}
\end{table}

\subsection{Zeroth and first order proposals (PMH0/1)}
PMH0 is referred to as a zero order (or marginal) proposal as it only makes use of the last accepted parameter to propose the new parameter. Essentially, this proposal is a Gaussian random walk scaled by a positive semi-definite (PSD) \textit{preconditioning matrix} $\mathcal{P}^{-1}$. The performance of PMH0 is highly dependent on $\mathcal{P}^{-1}$, which is tedious and difficult to estimate as it should be selected as the unknown posterior covariance, see \cite{SherlockThieryRobetsRosenthal2015}.

Furthermore, it is known that gradient information can be useful to give the proposal a mode-seeking behaviour. This can be beneficial both initially to find the mode and for increasing mixing by keeping the Markov chain in areas with high posterior probability. This information can be included by making use of noisy gradient ascent update, where $\widehat{\mathcal{G}}(\theta'|u')$ denotes the particle estimate of the gradient of the log-posterior given by $\mathcal{G}(\theta')=\nabla \log \pi(\theta) |_{\theta=\theta'}$. Again, we scale the step size and the gradient by $\mathcal{P}^{-1}$ resulting in the PMH1 proposal.

\subsection{Second order proposal (PMH2)}
An alternative is to make use of a noisy Newton update as the proposal by replacing $\mathcal{P}$ with $\widehat{\mathcal{H}}(\theta'|u')$, which denotes the particle estimate of the negative Hessian of the log-posterior given by $\mathcal{H}(\theta')= - \nabla^2 \log \pi(\theta) |_{\theta=\theta'}$. This results in the second order PMH2 proposal discussed by \cite{DahlinLindstenSchon2014c}, which relies on accurate estimates of the Hessian but these often require many particles and therefore incur a high computational cost. This problem is encountered in e.g.\ the ABC approximation of the $\alpha$-stable model in \eqref{eq:SSMdefABC}.

The new quasi-PMH2 (qPMH2) proposal circumvents this problem by constructing a local approximation of the Hessian based on a quasi-Newton update, which only makes use of gradient information. The update is inspired by the limited-memory BFGS algorithm \citep{Nocedal1980,NocedalWright2006} given by
\begin{align}
	&B^{-1}_{l+1}(\theta') 
	= 
	\big(
	\mathbf{I}_p - \rho_l s_l g_l^{\top}
	\big)
	B^{-1}_l
	\big(
	\mathbf{I}_p - \rho_l g_l s_l^{\top}
	\big)	
	+
	\rho_l s_l s_l^{\top},
	\label{eq:quasiNewtonUpdate}
\end{align}
with $\rho_l^{-1} = g_l^{\top} s_l$,  $s_l = \theta_{I(l)} - \theta_{I(l-1)}$ and $g_l = \widehat{\mathcal{G}}(\theta_{I(l)}|u_{I(l)}) - \widehat{\mathcal{G}}(\theta_{I(l-1)}|u_{I(l-1)})$. The update is iterated over $l \in \{1, 2, \ldots, M-1 \}$ with $I(l) = k - l$, i.e.\ over the $M-1$ previous states of the Markov chain. Hence, we refer to $M$ as the memory length of the proposal. We initialise the update with $ B_1^{-1} = \rho_1^{-1} ( g_1^{\top} g_1 )^{-1} \mathbf{I}_p$ and make use of a PMH0 proposal with $\mathcal{P}= \delta^{-1} \mathbf{I}_p$ for the first $M$ iterations, where $\delta > 0$ is defined by the user. The resulting estimate of the negative Hessian is given by $\widehat{\mathcal{H}}(\theta'|u') = -B_{M}(\theta')$. See Appendix~\ref{app:quasiNewton} for more details regarding the implementation of the qPMH2 proposal and the complete algorithm in Algorithm~\ref{alg:qnproposal}.

An apparent problem with using a quasi-Newton approximation of the Hessian is that the resulting proposal is no longer Markov (resulting in a non-standard MCMC). However, as shown by \cite{ZhangSutton2011}, it is still possible to obtain a valid algorithm by viewing the chain as an $M$-dimensional Markov chain. In effect, this amounts to using the sample at lag $M$ as the basis for the proposal. Hence, we set $\{\theta',u'\}=\{ \theta_{k-M}, u_{k-M}\}$ and $\epsilon_2=1$ in the PMH2 proposal in Table~\ref{tbl:pmh-proposals}. Furthermore, in case of a rejection we set $\{\theta_k, u_k\} = \{\theta_{k-M}, u_{k-M}\}$. We refer to \cite{ZhangSutton2011} for further details.

The approximate Hessian $\widehat{\mathcal{H}}(\theta'|u')$ has to be PSD to be a valid covariance matrix. This can be problematic when the Markov chain is located in areas with low posterior probability or sometimes due to noise in the gradient estimates. In our experience, this happens occasionally in the stationary regime, i.e.\ after the burn-in phase. However, when it happens $\widehat{\mathcal{H}}(\theta'|u')$ is corrected by the hybrid approach discussed by \cite{DahlinLindstenSchon2014c}.

\section{Proposal for auxiliary variables}
\label{sec:abc}
%To implement the PMH-ABC algorithm, we need estimates of the likelihood and the gradient.
% These can be obtained by the SMC-ABC algorithm in combination with a particle smoother.
To implement qPMH2, we require estimates of the likelihood and the gradient of the log-posterior. These are obtained by running SMC-ABC which corresponds to simulating the auxiliary variables $u$. In this section, we show how to estimate the likelihood and its gradient by the fixed-lag (FL; \citealp{KitagawaSato2001}) smoother.

\subsection{SMC-ABC algorithm}
\label{sec:abc:smc}
SMC-ABC \citep{JasraSinghMartinMcCoy2012} relies on a reformulation of the nonlinear SSM~\eqref{eq:SSMdef}. We start by perturbing the observations $y_t$ to obtain $\check{y}_{1:T}$ by
\begin{align}
	\check{y}_t = \psi( y_t ) + \epsilon z_t, \quad z_t \sim \rho_{\epsilon}, \quad \text{for } t=1,\ldots,T,
	\label{eq:ABCnoise}
\end{align}
where $\psi$ denotes a one-to-one transformation and $\rho_{\epsilon}$ denotes a kernel, e.g.\ Gaussian or uniform, with $\epsilon$ as the bandwidth or \textit{tolerance parameter}.  We continue with assuming that there exists some random variables $v_t \sim \nu_{\theta}(v_t|x_t)$ such that we can generate a sample from $g_{\theta}(y_t|x_t)$ by the transformation $y_t = \tau_{\theta}(v_t,x_t)$. An example is the \textit{Box-Muller transformation} to obtain a Gaussian random variable from two uniforms, see Appendix~\ref{app:impdetails}. 

To obtain the perturbed SSM, we introduce $\check{x}_t^{\top}=(x_t^{\top},v_t^{\top})$ as the new state variable with the dynamics
\begin{subequations}
\begin{align}
	\check{x}_{t+1} | \check{x}_t &\sim \Xi_{\theta}(\check{x}_{t+1}|\check{x}_t) = \nu_{\theta}(v_{t+1}|x_{t+1}) f_{\theta}(x_{t+1}|x_{t}),
\end{align}%
and the likelihood is modelled by
\begin{align}
	\check{y}_{t} | \check{x}_t & \sim h_{\theta,\epsilon}(\check{y}_{t} | \check{x}_t) 
	= \rho_{\epsilon} \big( \check{y}_t - \psi( \tau_{\theta}(\check{x}_t) ) \big),
\end{align}
\label{eq:SSMdefABC}%
\end{subequations}%
which follows from the perturbation in \eqref{eq:ABCnoise}. With this reformulation, we can construct SMC-ABC as outlined in Algorithm~\ref{alg:smc}, which is a standard SMC algorithm applied to the perturbed model. Note that, we do not require any evaluations of the intractable density $g_{\theta}(y_t|x_t)$. Instead, we only simulate from this distribution and compare the simulated and observed (perturbed) data by $\rho_{\epsilon}$. 

The accuracy of the ABC approximation is determined by $\epsilon$, where we recover the original formulation in the limit when $\epsilon \rightarrow 0$. In practice, this is not possible and we return to study the impact of a non-zero $\epsilon$ in Section~\ref{sec:results:lgss}.

\begin{algorithm}[!t]
\caption{\textsf{Sequential Monte Carlo with approximate Bayesian computations (SMC-ABC)}}
\textsc{Inputs:} $\check{y}_{1:T}$ (perturbed data), the SSM \eqref{eq:SSMdefABC}, $N \in \mathbb{N}$ (no.\ particles), $\epsilon > 0$ (tolerance parameter), $\Delta \in [0,T) \subset \mathbb{N}$ (lag). \\
\textsc{Outputs:} $\widehat{p}_{\theta}(\check{y}_{1:T}|u)$, $\widehat{\mathcal{G}}(\theta|u)$ (est.\ of likelihood and gradient). \\
\textsc{Note:} all operations are carried out over $i,j = 1, \ldots, N$.
\algrule[.4pt]
\begin{algorithmic}[1]
	\STATE Sample $\check{x}^{(i)}_0 \sim \mu_{\theta}(x_0) \nu_{\theta}(v_0|x_0)$ and set $w_0^{(i)}=1/N$.
	\FOR{$t=1$ to $T$}
		\STATE Resample the particles by sampling a new ancestor index $a^{(i)}_t$ from a multinomial distribution with $\mathbb{P} \big( a^{(i)}_t = j \big) = w^{(j)}_{t-1}$.
		\STATE Propagate the particles by sampling $\check{x}_t^{(i)} \sim \Xi_{\theta} \big( \check{x}_{t}^{(i)} \big| \check{x}_{t-1}^{a^{(i)}_t} \big)$ and extending the trajectory by $\check{x}_{0:t}^{(i)} = \big\{ \check{x}_{0:t-1}^{a_t^{(i)}}, \check{x}_{t}^{(i)} \big\}$.
		\STATE Calculate the particle weights by $\widetilde{w}^{(i)}_t = h_{\theta,\epsilon} \big( \check{y}_{t},\check{x}_{t}^{(i)} \big)$ which by normalisation (over $i$) gives $w^{(i)}_t$. 
	\ENDFOR
	\STATE Estimate $\widehat{p}_{\theta}(\check{y}_{1:T}|u)$ by \eqref{eq:likeEst} and $\widehat{\mathcal{G}}(\theta|u)$ by \eqref{eq:FisherScoreParticleApproximation}.
\end{algorithmic}
\label{alg:smc}
\end{algorithm}

\subsection{Estimation of the likelihood}
\label{sec:abc:loglikeEst}
From Section~\ref{sec:overview}, we require an unbiased estimate of the likelihood to compute the acceptance probability \eqref{eq:PMHaprob}. This can be achieved by using $u$ generated by SMC-ABC. In this case, the auxiliary variables $u \triangleq \{ \{\check{x}_{0:t}^{(i)}\}_{i=1}^N\}_{t=0}^T$ are the \textit{particle system} composed of all the particles and their trajectories. The resulting likelihood estimator is given by
\begin{align}
	\widehat{p}_{\theta}(y_{1:T} | u)
	=
	\prod_{t=1}^T \left[  \frac{1}{N} \sum_{i=1}^N \widetilde{w}_{t}^{(i)} \right],
	\label{eq:likeEst}
\end{align}
where the unnormalised particle weights $\widetilde{w}_{t}^{(i)}$ are deterministic functions of~$u$. This is an unbiased and $N$-consistent estimator for the likelihood in the \textit{perturbed} model. However, the perturbation itself introduces some bias and additional variance compared with the original unperturbed model \citep{DeanSinghJasraPeters2014}. The former can result in biased parameter estimates and the latter can result in poor mixing of the Markov chain. We return to study the impact on the mixing numerically in Section~\ref{sec:results:lgss}.

\subsection{Estimation of the gradient of the log-posterior}
\label{sec:abc:gradientEst}
We also require estimates of the gradient of the log-posterior given $u$ to implement the proposals introduced in Table~\ref{tbl:pmh-proposals}. In \citet{DahlinLindstenSchon2014c}, this is accomplished by using the FL smoother together with the \textit{Fisher identity}. However, this requires accurate evaluations of the gradient of $\log g_{\theta}(y_t|x_t)$ with respect to $\theta$. As discussed by \cite{YildirimSinghDeanJasra2014}, we can circumvent this problem by the reformulation of the SSM in \eqref{eq:SSMdefABC} if the gradient of $\tau_{\theta}(\check{x}_t)$ can be evaluated. This results in the gradient estimate
\begin{align}
	&\widehat{\mathcal{G}}(\theta'|u')
	=
	\nabla \log p(\theta) \Big|_{\theta=\theta'} +
	\sum_{t=1}^{T}
	\sum_{i=1}^N	
	w_{\kappa_t}\pIdx{i}
	\xi_{\theta'} \Big( \tilde{z}_{\kappa_t,t}\pIdx{i}, \tilde{z}_{\kappa_t,t-1}\pIdx{i} \Big),
	\label{eq:FisherScoreParticleApproximation}
	\\
	&\xi_{\theta'}(\check{x}_t,\check{x}_{t-1}) 
	 \triangleq 
	\nabla \log \Xi_{\theta}(\check{x}_t|\check{x}_{t-1}) \big|_{\theta=\theta'} + \nabla \log h_{\theta}(\check{y}_t|\check{x}_{t}) \big|_{\theta=\theta'}, \nonumber		
\end{align}
where $\tilde{z}_{\kappa_t,t}^{(i)}$ denotes the ancestor at time~$t$ of particle $\check{x}_{\kappa_t}^{(i)}$ and $\tilde{z}_{\kappa_t,t-1:t}^{(i)} = \{ \tilde{z}_{\kappa_t,t-1}^{(i)}, \tilde{z}_{\kappa_t,t}^{(i)} \}$. 

The estimator in \eqref{eq:FisherScoreParticleApproximation} relies on the assumption that the SSM is mixing quickly, which means that past states have a diminishing influence on future states and observations. More specifically, we assume that $p_{\theta}(x_t|y_{1:T}) \approx p_{\theta}(x_t|y_{1:\kappa_t})$, with $\kappa_t=\min\{t+\Delta,T\}$ and lag $\Delta \in [0,T) \subset \mathbb{N}$. Note that this estimator is biased, but this is compensated for by the accept-reject step in Algorithm~\ref{alg:pmh} and does not effect the stationary distribution of the Markov chain. See \citet{DahlinLindstenSchon2014c} for details. 

\section{Numerical illustrations}
\label{sec:results}
We evaluate qPMH2 by two illustrations with synthetic and real-world data. In the first model, we can evaluate $g_{\theta}(y_t|x_t)$ exactly in closed-form, which is useful to compare standard PMH and PMH-ABC. In the second model, the likelihood is intractable and therefore only PMH-ABC can be used. See Appendix~\ref{app:impdetails} for implementation details.

\begin{figure}[p]
	\centering
	\includegraphics[width=\columnwidth]{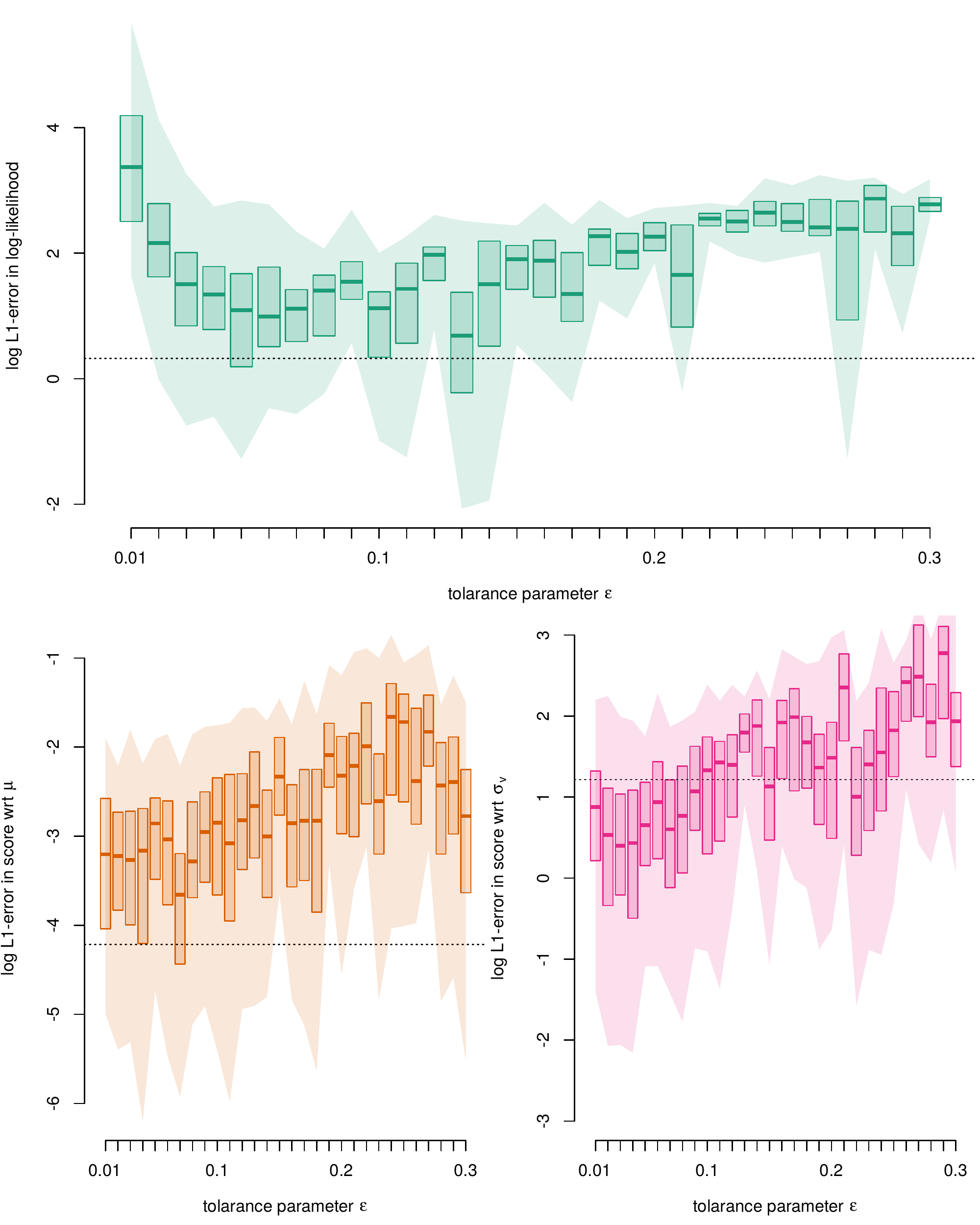}
	\caption{The log-$L_1$-error in the log-likelihood (top) and gradient with respect to $\mu$ (lower left) and $\sigma_v$ (lower right) using SMC-ABC when varying $\epsilon$. The shaded area and the dotted lines indicate maximum/minimum error and the standard SMC error, respectively. The plots are created from the output of $100$ Monte Carlo runs on a single synthetic data set.}
	\label{fig:lgss-epsilon}
\end{figure}

\subsection{Linear Gaussian SSM}
\label{sec:results:lgss}
Consider the following LGSS model
\begin{subequations}
\begin{align}
	x_{t+1} | x_t &\sim \mathcal{N} \Big( x_{t+1} ; \mu + \phi ( x_t - \mu ),\sigma_v^2 \Big), \\
	y_{t}   | x_t &\sim \mathcal{N} \Big( y_t; x_t, 0.1^2 \Big),
\end{align}%
\label{eq:lgssmodel}%
\end{subequations}%
\noindent with $\theta=\{\mu,\phi,\sigma_v\}$ and $\mu \in \mathbb{R}$, $\phi \in (-1,1)$ and $\sigma_v \in \mathbb{R}_{+}$. A synthetic data set consisting of a realisation with $T=250$ observations is simulated from the model using the parameters $\{0.2,0.8,1.0\}$. We begin by investigating the accuracy of Algorithm~\ref{alg:smc} for estimating the log-likelihood and the gradients of the log-posterior with respect to $\theta$. The error of these estimates are computed by comparing with the true values obtain by a Kalman smoother.

In Figure~\ref{fig:lgss-epsilon}, we present the log-$L_1$ error of the log-likelihood and the gradients for different values of $\epsilon$. The error in the gradient with respect to $\phi$ is not presented here, but is similar to the gradients for $\mu$.  We see that the error in both the log-likelihood and the gradient are minimized when $\epsilon \approx 0.05 - 0.10$. Here, SMC-ABC achieve almost the same error as standard SMC. However, when $\epsilon$ grows larger SMC-ABC suffers from an increasing bias resulting from a deteriorating approximation in~\eqref{eq:ABCnoise}. We conclude that this results in a bias in the parameter estimates as the bias in the log-likelihood estimate propagates to the parameter posterior estimate.

We now consider estimating the parameters in \eqref{eq:lgssmodel}. In this model, we can compare standard PMH with PMH-ABC to study the impact using ABC to approximate the log-likelihood. For this comparison, we quantify the mixing of the Markov chain using the estimated IF given by
\begin{align}
	\widehat{\textsf{IF}}(\theta_{K_b:K}) = 1 + 2 \sum_{l=1}^{L} \widehat{\rho}_l (\theta_{K_b:K}),
\end{align}
where $\widehat{\rho}_l(\theta_{K_b:K})$ denotes the empirical autocorrelation at lag $l$ of $\theta_{K_b:K}$, and $K_b$ is the \textit{burn-in} time. A small value of IF indicates that we obtain many uncorrelated samples from the target distribution. This implies that the chain is mixing well and that $\sigma^2_{\varphi}$ in \eqref{eq:functionalPosteriorCLT} is rather small. We make use of two different $L$ to compare the mixing: (i) the smallest $L$ such that $|\widehat{\rho}_L(\theta_{K_b:K})| < 2/\sqrt{K-K_b}$ (i.e.\ when statistical significant is lost) and (ii) a fixed $L=1,000$.

\begin{table}[t]
\caption{The IF values for the LGSS model \eqref{eq:lgssmodel}.}
\begin{center}
\begin{tabular}{llccccc}
\toprule
& & & \multicolumn{2}{c}{$\min \mathsf{IF}$} & \multicolumn{2}{c}{$\max \mathsf{IF}$} \\
\cmidrule(r){4-5} \cmidrule(r){6-7}
& Alg. & Acc. & Median & IQR & Median & IQR  \\
\midrule
\parbox[t]{2mm}{\multirow{7}{*}{\rotatebox[origin=c]{90}{$L$ adapted}}}
&PMH0      & 0.28 & 12.13 & 1.53  & 13.71 & 1.23 \\
&PMH1      & 0.78 & 11.28 & 0.50  & 14.50 & 1.45 \\
&qPMH2     & 0.55 & \textbf{3.00} & 0.03 & \textbf{3.01} & 0.07  \\
\cmidrule(r){2-7}
&PMH0-ABC  & 0.14 & 29.66 & 10.37 & 34.04 & 9.36 \\
&PMH1-ABC  & 0.31 & 33.09 & 5.45 & 38.32 & 14.42 \\
&qPMH2-ABC & 0.45 & \textbf{3.00} & 0.02 & \textbf{3.03} & 0.08  \\
\midrule
\parbox[t]{2mm}{\multirow{7}{*}{\rotatebox[origin=c]{90}{$L=1,000$}}}
&PMH0      & 0.28 & 7.96  & 9.60  & 10.92 & 6.00 \\
&PMH1      & 0.78 & 9.47 & 4.39  & 10.60 & 8.19 \\
&qPMH2     & 0.55 & \textbf{5.40} & 3.01 & \textbf{8.98} & 6.90  \\
\cmidrule(r){2-7}
&PMH0-ABC  & 0.14 & 12.91 & 8.86 & 35.68 & 3.22 \\
&PMH1-ABC  & 0.31 & 27.34 & 23.63 & 35.84 & 30.31 \\
&qPMH2-ABC & 0.45 & \textbf{6.65} & 5.14 & \textbf{10.96} & 6.71  \\
\bottomrule
\end{tabular}
\end{center}
\label{tbl:results-if}
\end{table}

In Table~\ref{tbl:results-if}, we present the minimum and maximum IFs as the median and interquartile range (IQR) computed using $10$ Monte Carlo runs over the same data set. We note the good performance of the qPMH2 proposal which achieves similar (or better) mixing compared to the pre-conditioned proposals. Remember that PMH0/1 are tuned to their optimal performance using tedious pilot runs, see \cite{SherlockThieryRobetsRosenthal2015} and \cite{NemethSherlockFearnhead2014}. We present some additional diagnostic plots for PMH0 and qPMH2 (without ABC) in Appendix~\ref{app:additionalresults}.

In our experience, the performance of qPMH2 seems to be connected with the variance of $\widehat{\mathcal{G}}(\theta'|u')$. This is similar to the existing theoretical results for PMH1 \citep{NemethSherlockFearnhead2014}. For the LGSS model, we can compute the gradient with a small variance and therefore qPMH2 performs well. However, this might not be the case for all models and the performance of qPMH2 thus depends on both the model and which particle smoother is applied.

Finally, note that using ABC results in a smaller acceptance rate and worse mixing. This problem can probably be mitigated by adjusting $\epsilon$ and $N$ as discussed by \cite{BornnPillaiSmithWoordward2014} and \cite{Jasra2015}.

\begin{figure}[p]
	\centering
	\includegraphics[width=\columnwidth]{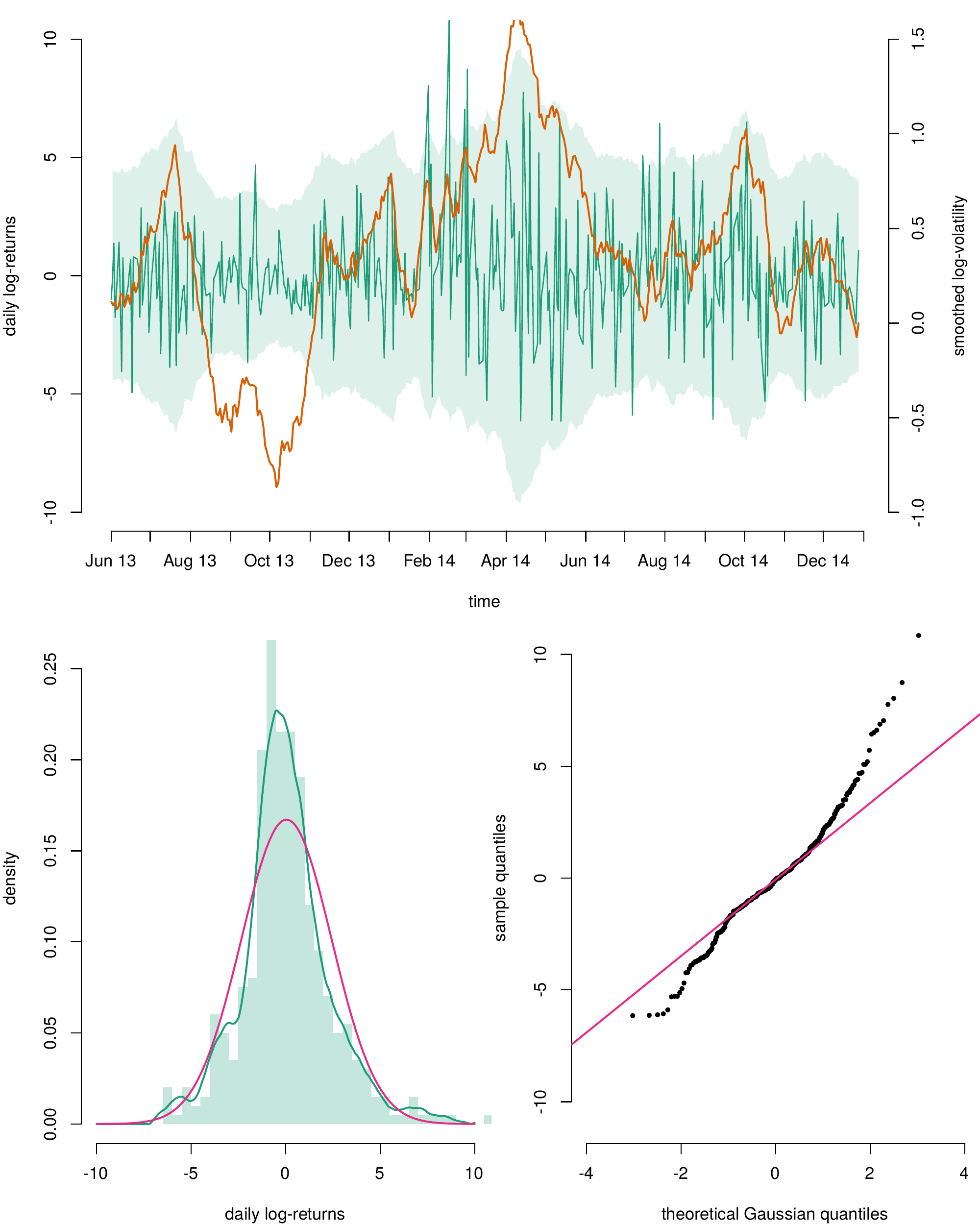}
	\caption{Upper: log-returns (green) and smoothed log-volatility (orange) of futures on coffee between June 1, 2013 and December 31, 2014. The shaded area indicate the approximate $95 \%$ credibility region for the log-returns estimated from the model. Lower left: kernel density estimate (green) and a Gaussian approximation (magenta). Lower right: QQ-plot comparing the quantiles of the data with the best Gaussian approximation (magenta).}
	\label{fig:pmh-sysid2015-aSVmodel}
\end{figure}

\begin{figure}[p]
	\centering
	\includegraphics[width=\columnwidth]{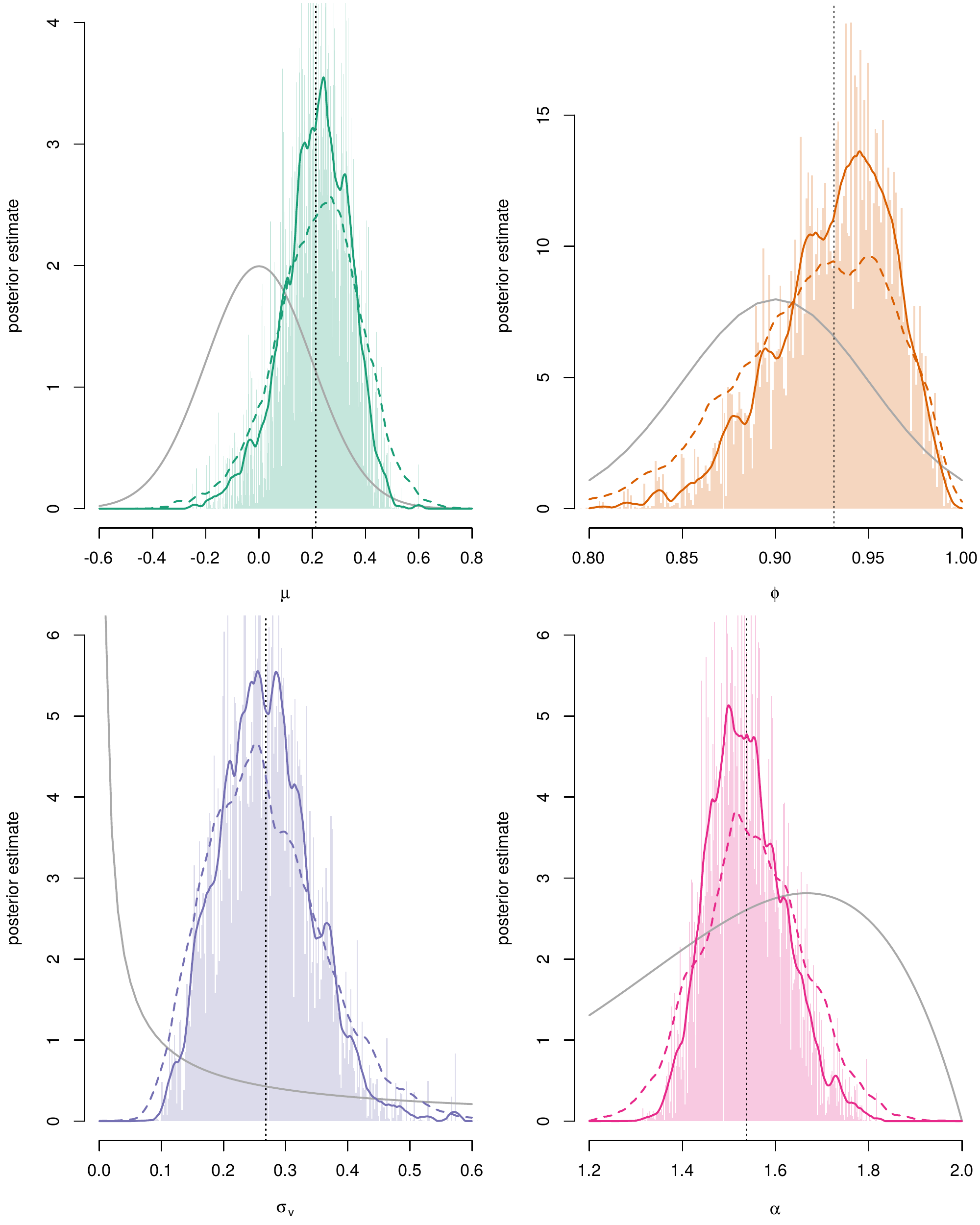}
	\caption{Parameter posteriors for \eqref{eq:aSVmodel} for $\mu$ (green), $\phi$ (orange), $\sigma_v$ (purple) and $\alpha$ (magenta) obtained by pooling the output from $10$ runs using qPMH2. The dashed lines indicate the corresponding estimates from PMH0. Dotted lines and grey densities indicate the estimate of the posterior means and the prior densities, respectively.}
	\label{fig:pmh-aSVmodel-posteriors}
\end{figure}

\subsection{Modelling the volatility in coffee futures}
\label{sec:results:coffee}
Consider the problem of modelling the volatility of the log-returns of future contracts on coffee using the $T=399$ observations in Figure~\ref{fig:pmh-sysid2015-aSVmodel}. A prominent feature in this type of data is jumps (present around March, 2014). These are typically the result of sudden changes in the market due to e.g.\ news arrivals and it has been proposed to make use of an $\alpha$-stable distribution to model these jumps, see e.g.\ \citet{LombardiCalzolari2009} and \citet{DahlinVillaniSchon2015}.

Therefore, we consider a stochastic volatility model with symmetric $\alpha$-stable returns ($\alpha$SV) given by
\begin{subequations}
\begin{align}
	x_{t+1} | x_t &\sim \mathcal{N} \Big( x_{t+1}; \mu + \phi ( x_t - \mu ),\sigma_v^2 \Big), \\
	y_{t}   | x_t &\sim \mathcal{A} \Big( y_{t}; \alpha, \exp(x_t) \Big),
\end{align}%
\label{eq:aSVmodel}%
\end{subequations}%
\noindent with $\theta=\{\mu,\phi,\sigma_v,\alpha\}$. Here, $\mathcal{A} ( \alpha, \eta )$ denotes a symmetric $\alpha$-stable distribution with \textit{stability parameter} $\alpha \in (0,2)$ and \textit{scale parameter} $\eta \in \mathbb{R}_{+}$. As previously discussed, we cannot evaluate $g_{\theta}(y_t|x_t)$ for this model but it is possible to simulate from it. See Appendices~\ref{app:impdetails} and \ref{app:astable} for more details regarding the $\alpha$-stable distribution and methods for simulating random variables.

In Figure~\ref{fig:pmh-aSVmodel-posteriors}, we present the posterior estimates obtained from qPMH2, which corresponds to the estimated parameter posterior mean $\widehat{\theta}=\{0.214,0.931,0.268,1.538\}$. This indicates a slowly varying log-volatility with heavy-tailed log-returns as the Cauchy and Gaussian distribution corresponds to $\alpha=1$ and $\alpha=2$, respectively. We also compare with the posterior estimates from PMH0 and conclude that they are similar in both location and scale. 

Finally, we present the smoothed estimate of the log-volatility using $\widehat{\theta}$ in Figure~\ref{fig:pmh-sysid2015-aSVmodel}. The estimate seems reasonable and tracks the periods with low volatility (around October, 2013) and high volatility (around May, 2014).

\section{Conclusions}
\label{sec:conclusions}
We have demonstrated that qPMH2 exhibits similar or improved mixing when compared with pre-conditioned proposals with/without the ABC approximation. The main advantage is that qPMH2 does not require extensive tuning of the step sizes in the proposal to achieve good mixing, which can be a problem in practice for PMH0/1. The user only needs to choose $\delta$ and $M$, which in our experience are simpler to tune. Finally, qPMH2 only requires gradient information, which are usually simpler to obtain than directly estimate the Hessian of the log-posterior.

In future work, it would be interesting to analyse the impact of the variance of the gradient estimates on the mixing of the Markov chain in qPMH2. In our experience, qPMH2 performs well when the variance is small or moderate. However when the variance increases, the mixing can be worse than for PMH0. This motivates further theoretical study and development of better particle smoothing techniques for gradient estimation to obtain gradient estimates with lower variance. 

Another extension of this work is to consider models where $p$ is considerable larger than discussed in this paper. This would probably lead to an even greater increase in mixing when using qPMH2 compared with PMH0. This effect is theoretically and empirically examined by \cite{NemethSherlockFearnhead2014}. In this context, it would also be interesting to implement a quasi-Newton proposal in a particle Hamiltonian Monte Carlo (HMC) algorithm in the spirit of \cite{ZhangSutton2011}. This as HMC algorithms are known to greatly increase mixing for some models when the gradient and log-likelihood can be evaluated analytically. However, no particle version of HMC has yet been proposed in the literature but the possibility is considered in the discussions following \cite{GirolamiCalderhead2011}.

\newpage
The source code and data for the LGSS model as well as some supplementary material are available from \url{https://github.com/compops/qpmh2-sysid2015/}.

\section*{Acknowledgements}
The simulations were performed on resources provided by the Swedish National Infrastructure for Computing (SNIC)  at Link\"{o}ping University, Sweden.

\clearpage
\bibliographystyle{plainnat}
\bibliography{dahlin}

\begin{thebibliography}{25}
\providecommand{\natexlab}[1]{#1}
\providecommand{\url}[1]{\texttt{#1}}
\expandafter\ifx\csname urlstyle\endcsname\relax
  \providecommand{\doi}[1]{doi: #1}\else
  \providecommand{\doi}{doi: \begingroup \urlstyle{rm}\Url}\fi

\bibitem[Andrieu and Roberts(2009)]{AndrieuRoberts2009}
C.~Andrieu and G.~O. Roberts.
\newblock {The pseudo-marginal approach for efficient Monte Carlo
  computations}.
\newblock \emph{The Annals of Statistics}, 37\penalty0 (2):\penalty0 697--725,
  2009.

\bibitem[Andrieu et~al.(2010)Andrieu, Doucet, and
  Holenstein]{AndrieuDoucetHolenstein2010}
C.~Andrieu, A.~Doucet, and R.~Holenstein.
\newblock {Particle Markov chain Monte Carlo methods}.
\newblock \emph{Journal of the Royal Statistical Society: Series B (Statistical
  Methodology)}, 72\penalty0 (3):\penalty0 269--342, 2010.

\bibitem[Bornn et~al.(2014)Bornn, Pillai, Smith, and
  Woordward]{BornnPillaiSmithWoordward2014}
L.~Bornn, N.~Pillai, A.~Smith, and D.~Woordward.
\newblock {A Pseudo-Marginal Perspective on the ABC Algorithm}.
\newblock \emph{Pre-print}, 2014.
\newblock arXiv:1404.6298v1.

\bibitem[Chambers et~al.(1976)Chambers, Mallows, and
  Stuck]{ChambersMallowsStuck1976}
J.~M. Chambers, C.~L. Mallows, and B.~Stuck.
\newblock A method for simulating stable random variables.
\newblock \emph{{Journal of the American Statistical Association}}, 71\penalty0
  (354):\penalty0 340--344, 1976.

\bibitem[Dahlin et~al.(2015{\natexlab{a}})Dahlin, Lindsten, and
  Sch\"{o}n]{DahlinLindstenSchon2014c}
J.~Dahlin, F.~Lindsten, and T.~B. Sch\"{o}n.
\newblock {Particle Metropolis-Hastings using gradient and Hessian
  information}.
\newblock \emph{Statistics and Computing}, 25\penalty0 (1):\penalty0 81--92,
  2015{\natexlab{a}}.

\bibitem[Dahlin et~al.(2015{\natexlab{b}})Dahlin, Lindsten, and
  Sch\"{o}n]{DahlinLindstenSchon2015a}
J.~Dahlin, F.~Lindsten, and T.~B. Sch\"{o}n.
\newblock {Particle Metropolis-Hastings using gradient and Hessian
  information}.
\newblock \emph{Statistics and Computing}, 25\penalty0 (1):\penalty0 81--92,
  2015{\natexlab{b}}.

\bibitem[Dahlin et~al.(2015{\natexlab{c}})Dahlin, Villani, and
  Sch\"{o}n]{DahlinVillaniSchon2015}
J.~Dahlin, M.~Villani, and T.~B. Sch\"{o}n.
\newblock {Efficient approximate Bayesian inference for models with intractable
  likelihoods}.
\newblock \emph{Pre-print}, 2015{\natexlab{c}}.
\newblock arXiv:1506.06975v1.

\bibitem[Dean et~al.(2014)Dean, Singh, Jasra, and
  Peters]{DeanSinghJasraPeters2014}
T.~A. Dean, S.~S. Singh, A.~Jasra, and G.~W. Peters.
\newblock {Parameter estimation for hidden Markov models with intractable
  likelihoods}.
\newblock \emph{Scandinavian Journal of Statistics}, 41\penalty0 (4):\penalty0
  970--987, 2014.

\bibitem[Doucet and Johansen(2011)]{DoucetJohansen2011}
A.~Doucet and A.~Johansen.
\newblock A tutorial on particle filtering and smoothing: Fifteen years later.
\newblock In D.~Crisan and B.~Rozovsky, editors, \emph{The Oxford Handbook of
  Nonlinear Filtering}. Oxford University Press, 2011.

\bibitem[Girolami and Calderhead(2011)]{GirolamiCalderhead2011}
M.~Girolami and B.~Calderhead.
\newblock {Riemann manifold Langevin and Hamiltonian Monte Carlo methods}.
\newblock \emph{Journal of the Royal Statistical Society: Series {B}
  (Statistical Methodology)}, 73\penalty0 (2):\penalty0 1--37, 2011.

\bibitem[Jasra(2015)]{Jasra2015}
A.~Jasra.
\newblock {Approximate Bayesian Computation for a Class of Time Series Models}.
\newblock \emph{International Statistical Review}, \penalty0 (accepted for
  publication), 2015.

\bibitem[Jasra et~al.(2012)Jasra, Singh, Martin, and
  McCoy]{JasraSinghMartinMcCoy2012}
A.~Jasra, S.~S. Singh, J.~S. Martin, and E.~McCoy.
\newblock {Filtering via approximate Bayesian computation}.
\newblock \emph{Statistics and Computing}, 22\penalty0 (6):\penalty0
  1223--1237, 2012.

\bibitem[Kitagawa and Sato(2001)]{KitagawaSato2001}
G.~Kitagawa and S.~Sato.
\newblock {Monte Carlo smoothing and self-organising state-space model}.
\newblock In A.~Doucet, N.~de~Fretias, and N.~Gordon, editors, \emph{Sequential
  Monte Carlo methods in practice}, pages 177--195. Springer, 2001.

\bibitem[Lombardi and Calzolari(2009)]{LombardiCalzolari2009}
M.~J. Lombardi and G.~Calzolari.
\newblock Indirect estimation of $\alpha$-stable stochastic volatility models.
\newblock \emph{Computational Statistics \& Data Analysis}, 53\penalty0
  (6):\penalty0 2298--2308, 2009.

\bibitem[Marin et~al.(2012)Marin, Pudlo, Robert, and
  Ryder]{MarinPudloRobertRyder2012}
J-M. Marin, P.~Pudlo, C.~P. Robert, and R.~J. Ryder.
\newblock {Approximate Bayesian computational methods}.
\newblock \emph{Statistics and Computing}, 22\penalty0 (6):\penalty0
  1167--1180, 2012.

\bibitem[Meyn and Tweedie(2009)]{MeynTweedie2009}
S.~P. Meyn and R.~L. Tweedie.
\newblock \emph{Markov chains and stochastic stability}.
\newblock Cambridge University Press, 2009.

\bibitem[Nemeth et~al.(2014)Nemeth, Sherlock, and
  Fearnhead]{NemethSherlockFearnhead2014}
C.~Nemeth, C.~Sherlock, and P.~Fearnhead.
\newblock {Particle Metropolis adjusted Langevin algorithms}.
\newblock \emph{Pre-print}, 2014.
\newblock arXiv:1412.7299v1.

\bibitem[Nocedal(1980)]{Nocedal1980}
J.~Nocedal.
\newblock {Updating quasi-Newton matrices with limited storage}.
\newblock \emph{Mathematics of Computation}, 35\penalty0 (151):\penalty0
  773--782, 1980.

\bibitem[Nocedal and Wright(2006)]{NocedalWright2006}
J.~Nocedal and S.~Wright.
\newblock \emph{{Numerical Optimization}}.
\newblock Springer, 2 edition, 2006.

\bibitem[Nolan(2003)]{Nolan2003}
J.~Nolan.
\newblock \emph{Stable distributions: models for heavy-tailed data}.
\newblock Birkhauser, 2003.

\bibitem[Peters et~al.(2012)Peters, Sisson, and Fan]{PetersSissonFan2012}
G.~W. Peters, S.~A. Sisson, and Y.~Fan.
\newblock {Likelihood-free Bayesian inference for $\alpha$-stable models}.
\newblock \emph{Comput. Stat. Data Anal.}, 56\penalty0 (11):\penalty0
  3743--3756, November 2012.

\bibitem[Schraudolph et~al.(2007)Schraudolph, Yu, and
  G{\"u}nter]{SchraudolphYuGunter2007}
N.~Schraudolph, J.~Yu, and S.~G{\"u}nter.
\newblock {A stochastic quasi-Newton method for online convex optimization}.
\newblock In \emph{Proceedings of the 11th International Conference on
  Artificial Intelligence and Statistics}, San Juan, Puerto Rico, mar 2007.

\bibitem[Sherlock et~al.(2015)Sherlock, Thiery, Roberts, and
  Rosenthal]{SherlockThieryRobetsRosenthal2015}
C.~Sherlock, A.~H. Thiery, G.~O. Roberts, and J.~S. Rosenthal.
\newblock {On the efficency of pseudo-marginal random walk Metropolis
  algorithms}.
\newblock \emph{The Annals of Statistics}, 43\penalty0 (1):\penalty0 238--275,
  2015.

\bibitem[Y{\i}ld{\i}r{\i}m et~al.(2014)Y{\i}ld{\i}r{\i}m, Singh, Dean, and
  Jasra]{YildirimSinghDeanJasra2014}
S.~Y{\i}ld{\i}r{\i}m, S.~S. Singh, T.~Dean, and A.~Jasra.
\newblock {Parameter estimation in hidden Markov models with intractable
  likelihoods using sequential Monte Carlo}.
\newblock \emph{Journal of Computational and Graphical Statistics}, \penalty0
  (accepted for publication), 2014.

\bibitem[Zhang and Sutton(2011)]{ZhangSutton2011}
Y.~Zhang and C.~A. Sutton.
\newblock {Quasi-Newton methods for Markov chain Monte Carlo}.
\newblock In J.~Shawe-Taylor, R.~S. Zemel, P.~L. Bartlett, F.~Pereira, and
  K.~Q. Weinberger, editors, \emph{Advances in Neural Information Processing
  Systems 24}, pages 2393--2401. 2011.

\end{thebibliography}

\clearpage
\appendix
\section{Implementation details}
\label{app:impdetails}
In Algorithm~\ref{alg:smc} for the LGSS model, we use a fully adapted SMC algorithm with $N=50$ and SMC-ABC with $N=2,500$, lag $\Delta=12$, $\epsilon=0.10$ and $\rho_{\epsilon}$ as the Gaussian density with standard deviation $\epsilon$. For the $\alpha$SV, we use the same settings expect for $N=5,000$. For qPMH2, we use the memory length $M=100$, $\delta=1,000$ and $n_{\text{hyb}} = 2,500$ samples for the hybrid method. We use $K=15,000$ iterations (discarding the first $K_b=5,000$ as burn-in) for all PMH algorithms and initialise in the maximum a posteriori (MAP) estimate obtained accordingly to \cite{DahlinVillaniSchon2015}. 

The pre-conditioning matrix $\mathcal{P}$ is estimated by pilot runs using PMH0, with step sizes based on the Hessian estimate obtained in the MAP estimation. The final step sizes are given by the rules of thumb by \citet{SherlockThieryRobetsRosenthal2015} and \citet{NemethSherlockFearnhead2014}, i.e.\ 
\begin{align*}
	\epsilon_0^2 = 2.562^2 p^{-1}, \qquad
	\epsilon_1^2 = 1.125^2 p^{-1/3}.
\end{align*}
Finally, we use the following prior densities
\begin{alignat*}{4}
	&p(\mu)      &&\sim \mathcal{TN}_{(0,1)}(\mu;0,0.2^2), \qquad
	&p(\phi)      &\sim \mathcal{TN}_{(-1,1)}(\phi;0.9,0.05^2), \\
	&p(\sigma_v) &&\sim \mathcal{G}(\sigma_v;0.2,0.2), \qquad
	&p(\alpha)    &\sim \mathcal{B}(\alpha/2;6,2),
\end{alignat*}
\noindent where $\mathcal{TN}_{(a,b)}(\cdot)$ denotes a truncated Gaussian distribution on $[a,b]$, $\mathcal{G}(a,b)$ denotes the Gamma distribution with mean $a/b$ and $\mathcal{B}(a,b)$ denotes the Beta distribution.

For SMC-ABC, we require the transformation $\tau(\check{x}_t)$ to simulate random variables from the two models. For the LGSS model, we use the identity transformation $\psi(x)=x$ and the Box-Muller transformation to simulate $y_t$ by
\begin{align*}
	\tau_{\theta}(\check{x}_t) = x_t + \sigma_e \sqrt{ - 2 \log v_{t,1} } \cos ( 2 \pi v_{t,2}),
\end{align*}
where $\{v_{t,1},v_{t,2}\} \sim \mathcal{U}[0,1]$. 

For the $\alpha$SV model, we use $\psi(x)=\arctan(x)$ as proposed by \citet{YildirimSinghDeanJasra2014} to make the variance in the gradient estimate finite. We generate samples from $\mathcal{A}(\alpha,\exp(x_t))$ for $\alpha \neq 1$ by
\begin{align*}
\tau_{\theta}(\check{x}_t) = \exp(x_t/2)
		\frac{ \sin (\alpha v_{t,2} ) }
		{[ \cos (v_{t,2}) ]^{1/\alpha}}
		\left[
		\frac
		{\cos \left[ (\alpha - 1) v_{t,2} \right]}{v_{t,1}}
		\right]^{\frac{1-\alpha}{\alpha}},
\end{align*}
where $\{ v_{t,1},v_{t,2} \} \sim \{ \textsf{Exp}(1),\mathcal{U}(-\pi/2,\pi/2) \}$. The real-world data in the $\alpha$SV model is computed as $y_t = 100 [ \log(s_{t}) - \log(s_{t-1}) ]$, where $s_t$ denotes the price of a future contract on coffee obtained from \url{https://www.quandl.com/CHRIS/ICE_KC2}.

\section{Implementation details for quasi-Newton proposal}
\label{app:quasiNewton}
%Theory. Implementation algorithm.
The quasi-Newton proposal adapts the Hessian estimate at iteration $k$ using the previous $M$ states of the Markov chain. Hence, the current proposed parameter depends on the previous $M$ states and therefore this can be seen as a Markov chain of order $M$. Following \cite{ZhangSutton2011}, we can analyse this chain as a first-order Markov chain on an extended $M$-fold product space $\Theta^M$. This results in that the stationary distribution can be written as 
\begin{align*}
	\bm{\pi}(\theta_{1:M}) = \prod_{i=1}^M \pi(\theta_{i}),
\end{align*}
where $\pi(\theta_{k,i})$ is defined as in \eqref{eq:parameterposterior}. To proceed with the analysis, we introduce the notation $\psi_k=\{\theta_k,u_k\}$ and $\psi_{k,1:M \setminus i}$ for the vector $\psi_{k,1:M}$ with $\psi_{k,i}$ removed for brevity. We can then update a component of $\psi_{k,1:M}$ using some transition kernel $T_i$ defined by
\begin{align*}
	T_i \Big( \psi_i, \psi_i' \big| \psi_{k,1:M \setminus i} \Big) = \delta \Big( \psi_{k,1:M \setminus i}, \psi'_{k,1:M \setminus i} \Big) B \Big( \psi_i, \psi_i' \big| \psi_{k,1:M \setminus i} \Big),
\end{align*}
where $B \Big( \psi_i, \psi_i' \big| \psi_{k,1:M \setminus i} \Big)$ denotes the quasi-Newton PMH2 proposal adapted using the last $M-1$ samples $\psi_{k,1:M \setminus i}$, analogously to \eqref{eq:quasiNewtonUpdate}. This is similar to a Gibbs-type step in which we update a component conditional on the remaining $M-1$ components. Here, this conditioning is used to construct the local Hessian approximation using the information in the remaining components. 

As this update leaves $\pi(\theta_{k,i})$ invariant, it follows that
\begin{align*}
	T \big( \psi_{k,1:M},\psi_{k,1:M}' \big) = T_1 \circ T_2 \circ \cdots \circ T_M \big( \psi_{k,1:M},\psi_{k,1:M}' \big),
\end{align*}
leaves $\bm{\pi}(\theta_{k,1:M})$ invariant. Hence, we can update all components of the $M$-dimensional Markov chain during each iteration $k$ of the sampler.

To implement the algorithm, we instead interpret the proposal as depending on the last $M-1$ states of the Markov chain. This results in a sliding window of samples, which we use to adapt the proposal. This results in two minor differences from a standard PMH algorithm. The first is that we center the proposal around the position of the Markov chain at $k-M$, i.e.\
\begin{align}
	q( \theta' | \psi_{k-M+1:k-1} ) = \mathcal{N} \big( \theta'; \theta_{k-M}, \Sigma_{\text{BFGS}}(\psi_{k-M+1:k-1} ) \big),
	\label{eq:qnproposal}
\end{align}
where $\Sigma_{\text{BFGS}}(\psi_{k-M+1:k-1} ) = -B_M(\theta')$ obtain by iterating \eqref{eq:quasiNewtonUpdate}. The second difference is that we set $\{\theta_k,u_k\} \leftarrow \{\theta_{k-M},u_{k-M}\}$ is the candidate parameter $\theta'$ is rejected. The complete procedure for proposing from $q( \theta'_k | \psi_{k-M+1:k-1} )$ is presented in Algorithm~\ref{alg:qnproposal}.

\begin{algorithm}[!t]
\caption{\textsf{Quasi-Newton proposal}}
\textsc{Inputs:} $\{\theta_{k-M:k-1},u_{k-M:k-1}\}$ (last $M$ states of the Markov chain), $\delta > 0$ (initial Hessian). \\
\textsc{Outputs:} $\theta'$ (proposed parameter).
\algrule[.4pt]
\begin{algorithmic}[1]
	\STATE Extract the $M^{\star}$ unique elements from $\{\theta_{k-M+1:k-1},u_{k-M+1:k-1}\}$ and sort them in ascending order (with respect to the log-likelihood) to obtain $\{\theta^{\star},u^{\star}\}$.
	\IF{ $M^{\star} \geq 2$ }
	\STATE Calculate $s_l$ and $y_l$ for $l=1,\ldots,M-2$ using the ordered pairs in $\{\theta^{\star},u^{\star}\}$ and their corresponding gradient estimates.
	\STATE Initialise the Hessian estimate $B_1^{-1} = \rho_1^{-1} (y_1^{\top} y_1)^{-1} \mathbf{I}_p$.
	\FOR{$l=1$ to $M^{\star}-1$}
		%\STATE Compute the difference in the parameters $s_l = \theta^{\star}_{l+1} - \theta_{l}$ and gradient $g_l = \widehat{\mathcal{G}}(\theta_{l+1}|u_{l+1}) - \widehat{\mathcal{G}}(\theta_{l}|u_{l})$ using \eqref{eq:FisherScoreParticleApproximation}.
		\STATE Carry out the update \eqref{eq:quasiNewtonUpdate} to obtain $B_{l+1}^{-1}$.
	\ENDFOR
	\STATE Set $\Sigma_{\text{BFGS}}(\psi_{k-M+1:k-1} ) = -B_{M^{\star}}(\theta')$.
	\ELSE
	\STATE Set $\Sigma_{\text{BFGS}}(\psi_{k-M+1:k-1} ) = \delta \mathbf{I}_p$.
	\ENDIF
	\STATE Sample from \eqref{eq:qnproposal} to obtain $\theta'$.
\end{algorithmic}
\label{alg:qnproposal}
\end{algorithm}

Some possible extensions to this noisy quasi-Newton inspired update are discussed by \cite{SchraudolphYuGunter2007} and \cite{ZhangSutton2011}. These include how to rearrange the update such that the Hessian estimate always is a PSD matrix. In this paper, we instead make use of the hybrid method discussed by \cite{DahlinLindstenSchon2015a} to handle these situations. In \cite{SchraudolphYuGunter2007}, the authors also discuss possible alternations to handle noisy gradients, which could be useful in some situations. In our experience, these alternations do not always improve the quality of the Hessian estimate as the noisy is stochastic and not the result of a growing data set as in the aforementioned paper. 

\section{Additional results}
\label{app:additionalresults}
\begin{figure}[p]
	\centering
	\includegraphics[width=\textwidth]{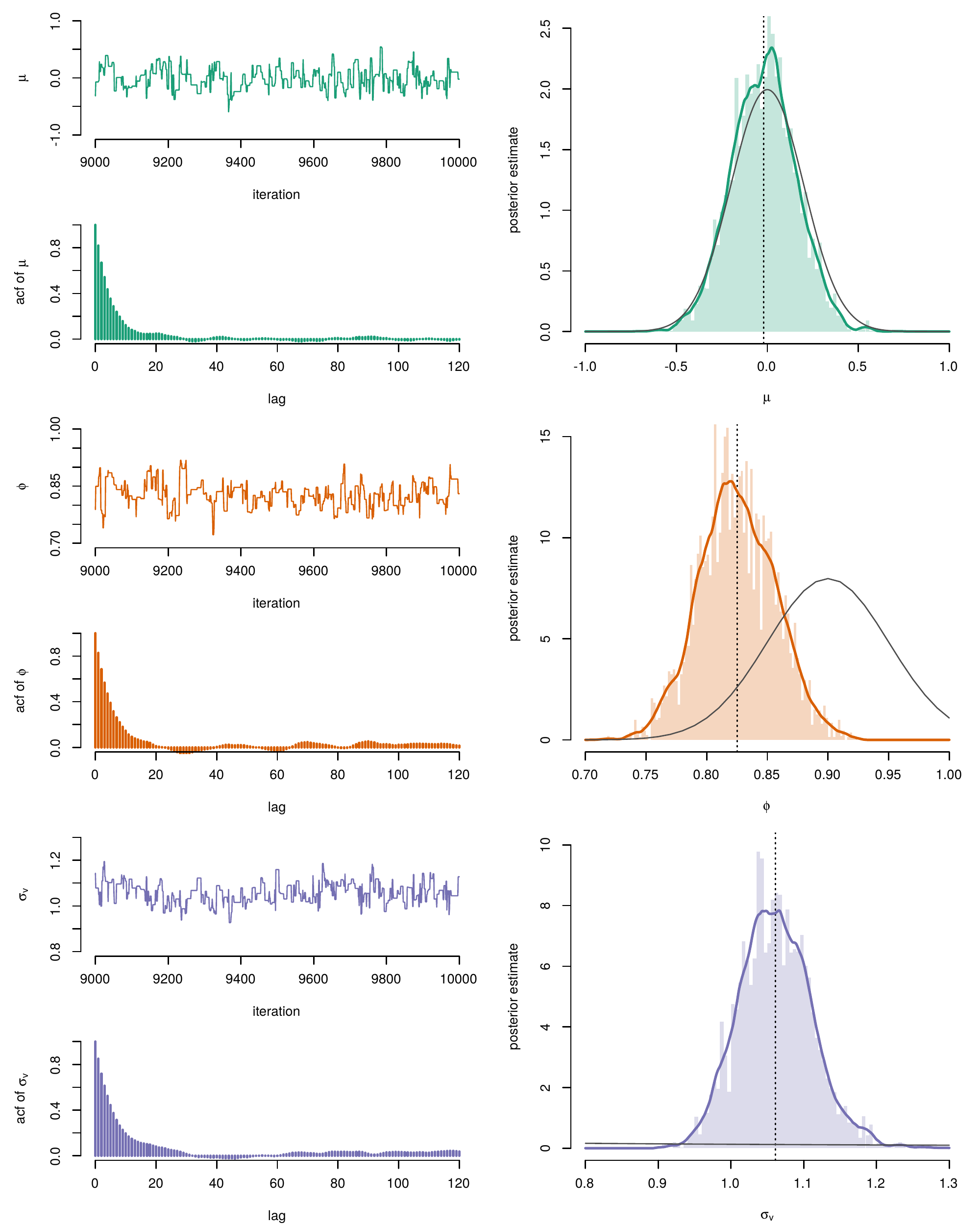}
	\caption{Trace plots and ACF estimates (left) for $\mu$ (green), $\phi$ (orange) and $\sigma_v$ (purple) in the LGSS model using pPMH0-ABC. The resulting posterior estimates (right) are also presented as histograms and kernel density estimates. The dotted vertical lines in the estimates of the posterior indicate its estimated mean. The grey lines indicate the parameter prior distributions.}
	\label{fig:example2-diagnosticplots-ppmh0}
\end{figure}

\begin{figure}[p]
	\centering
	\includegraphics[width=\textwidth]{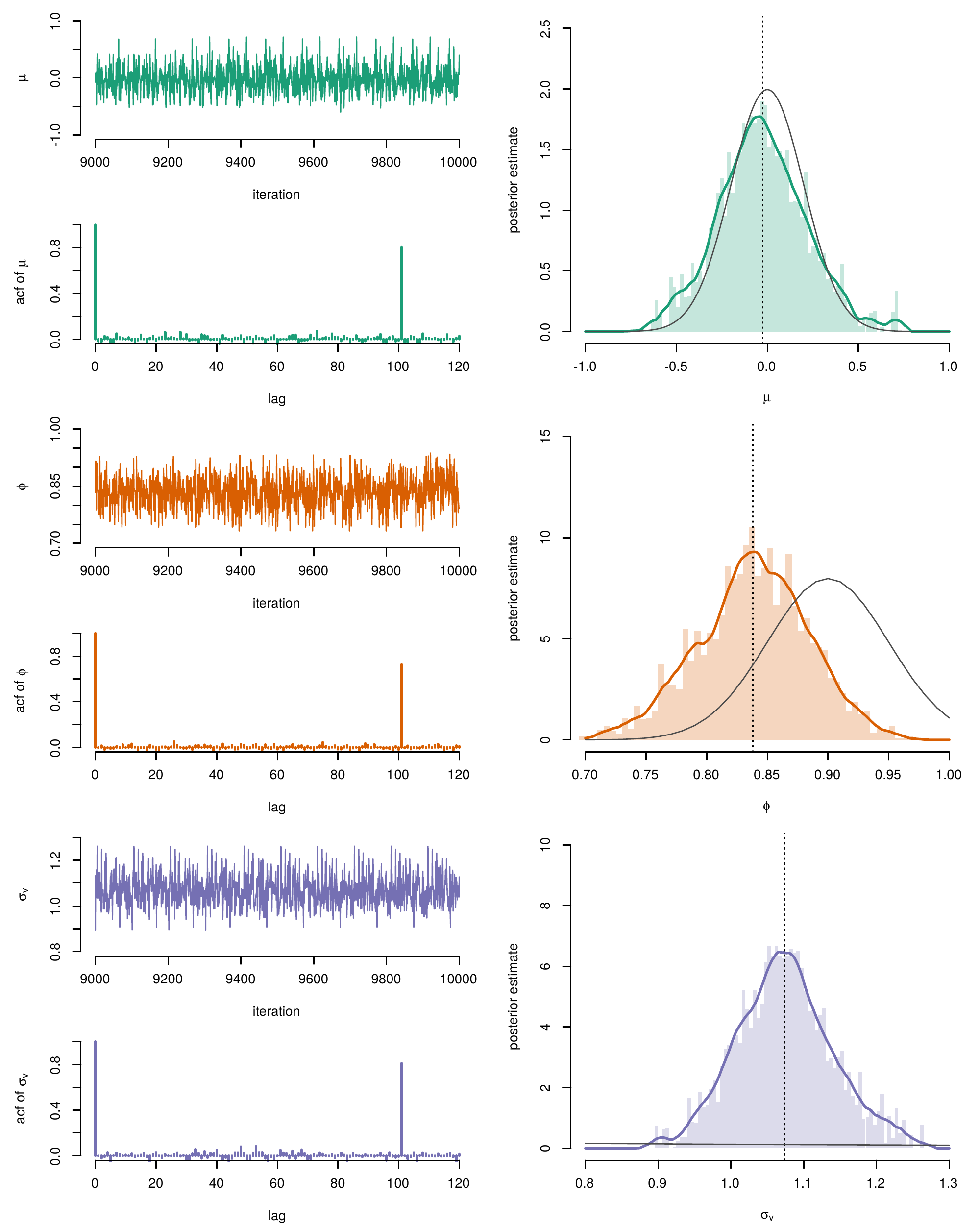}
	\caption{Trace plots and ACF estimates (left) for $\mu$ (green), $\phi$ (orange) and $\sigma_v$ (purple) in the LGSS model using qPMH2-ABC. The resulting posterior estimates (right) are also presented as histograms and kernel density estimates. The dotted vertical lines in the estimates of the posterior indicate its estimated mean. The grey lines indicate the parameter prior distributions.}
	\label{fig:example2-diagnosticplots-qpmh2}
\end{figure}

%Plots for trace and posteriors in LGSS.
In this section, we present some additional plots for the LGSS example in Section~\ref{sec:results:lgss} using PMH0 in Figure~\ref{fig:example2-diagnosticplots-ppmh0} and qPMH2 in Figure~\ref{fig:example2-diagnosticplots-qpmh2}. 

We note that the mixing is better for qPMH2 as the ACF decreases quicker as the lag increased compared with PMH0. However due to the quasi-Newton proposal, we obtain a large correlation coefficient at lag $M$. This is also reflected in the trace plots, which exhibits a periodic behaviour. Hence, we conclude that the mixing is increased in some sense when using qPMH2 compared with PMH0. The exact magnitude of this improvement depends on the method to compute the IF values.

\section{$\alpha$-stable distributions}
\label{app:astable}
This appendix summarises some important results regarding $\alpha$-stable distributions. For a more detailed presentation, see \cite{Nolan2003}, \cite{PetersSissonFan2012} and references therein.

\newpage
\begin{definition}[$\alpha$-stable distribution \citep{Nolan2003}]
An univariate $\alpha$-stable distribution denoted by $\mathcal{A}(\alpha,\beta,c,\mu)$ has the characteristic function
\begin{align*}
	\phi(t;\alpha,\beta,c,\mu)
	=
	\begin{cases}
	\exp \left\{ it \mu - |ct|^{\alpha} \left[ 1 - i \beta \tan \left( \frac{\pi \alpha}{2} \right) \textsf{sgn}(t) \right] \right\} & \text{ if } \alpha \neq 1, \\
	\exp \left\{ it \mu - |ct|^{\alpha} \left[ 1 + \frac{2 i \beta}{\pi} \textsf{sgn}(t) \ln |t| \right] \right\} & \text{ if } \alpha = 1,
	\end{cases}
\end{align*}
where $\alpha \in [0, 2]$ denotes the stability parameter, $\beta \in [-1, 1]$ denotes the skewness parameters, $c \in \mathbb{R}_{+}$ denotes the scale parameter and $\mu \in \mathbb{R}$ denotes the location parameter.
\label{def:alphastable1}
\end{definition}

The $\alpha$-stable distribution is typically defined through its characteristic function given in Definition~\ref{def:alphastable1}. Except for some special cases, we cannot recover the probability distribution function (pdf) from $\phi(t;\alpha,\beta,c,\mu)$ as it cannot be computed analytically. These exceptions are; (i) the Gaussian distribution $\mathcal{N}(\mu,2c^2)$ is recovered when $\alpha=2$ for any $\beta$ (as the $\alpha$-stable distribution is symmetric for this choice of $\alpha$), (ii) the Cauchy distribution $\mathcal{C}(\mu,c)$ is recovered when $\alpha=1$ and $\beta=0$, and (iii) the L\'{e}vy distribution $\mathcal{L}(\eta,c)$ is recovered when $\alpha=0.5$ and $\beta=1$. 

For all other choices of $\alpha$ and $\beta$, we cannot recover the pdf from $\phi(t;\alpha,\beta,c,\mu)$ due to the analytical intractability of the Fourier transform. However, we can often approximate the pdf using numerical methods as discussed in \cite{PetersSissonFan2012}. In this work, we make use of another approach based on ABC approximations to circumvent the intractability of the pdf. For this, we require to be able to sample from $\mathcal{A}(\alpha,\beta,c,\mu)$ for any parameters. An procedure for this discussed by \cite{ChambersMallowsStuck1976} is presented in Proposition~\ref{prop:simulate1}. 

\begin{proposition}[Simulating $\alpha$-stable variable \citep{ChambersMallowsStuck1976}] Assume that we can simulate $w \sim \textsf{Exp}(1)$ and $u \sim \mathcal{U}(-\pi/2,\pi/2)$. Then, we can obtain a sample from $\mathcal{A}(\alpha,\beta,1,0)$ by
\begin{align}
\bar{y} = 
	\begin{cases}
		\frac{ \sin \left[ \alpha ( u + T_{\alpha,\beta} ) \right] }
		{(\cos(\alpha T_{\alpha,\beta}) \cos (u) )^{1/\alpha}}
		\left[
		\frac
		{\cos \left[ \alpha T_{\alpha,\beta} + (\alpha - 1) u \right]}{w}
		\right]^{\frac{1-\alpha}{\alpha}} & \text{ if } \alpha \neq 1, \\[.9em]
		\frac{2}{\pi}
		\left[
		\left(
		\frac{\pi}{2}
		+ \beta u
		\right)
		\tan (u)
		-
		\beta
		\log 
		\frac{\frac{\pi}{2} w \cos u }
		{\frac{\pi}{2} + \beta u}
		\right]  & \text{ if } \alpha = 1,
	\end{cases}
\end{align}
where we have introduced the following notation
\begin{align*}
	T_{\alpha,\beta} &= \frac{1}{\alpha} \arctan \left( \beta \tan \left( \frac{\pi \alpha}{2} \right) \right).
\end{align*}
A sample from $\mathcal{A}(\alpha,\beta,c,\mu)$ is obtained by the transformation
\begin{align*}
	y = 
	\begin{cases}
		c \bar{y} + \mu & \text{ if } \alpha \neq 1, \\
		c \bar{y} + \left( \mu + \beta \frac{2}{\pi} c \log c \right) & \text{ if } \alpha = 1.
	\end{cases}
\end{align*}
\label{prop:simulate1}
\end{proposition}
  
\end{document}